\documentclass[aps,nofootinbib,showpacs,twocolumn]{revtex4-1}
\usepackage{graphicx}
\usepackage{dcolumn}
\usepackage{amsmath}
\usepackage{latexsym}
\usepackage{mathrsfs}
\usepackage[usenames,dvipsnames]{color}

\usepackage{hyperref}
\newcommand{\be}{\begin{equation}} 
\newcommand{\ee}{\end{equation}} 
\newcommand{\bea}{\begin{eqnarray}} 
\newcommand{\eea}{\end{eqnarray}} 

\newcommand{\eps}{\epsilon}
\begin{document}
%\title{Percolation of point configurations with hyperuniform  or GNF number-fluctuations}
\title{Percolation of systems having  hyperuniformity  or giant number-fluctuations}

\author{Sayantan Mitra$^1$}
\email{sayantan.pdf@iiserkol.ac.in }
\author{Indranil Mukherjee$^2$}
\email{indranil.mukherjee@icts.res.in}

\author{P. K. Mohanty$^1$}
\email{pkmohanty@iiserkol.ac.in}
\affiliation {$^1$Department of Physical Sciences, Indian Institute of Science Education and Research Kolkata, Mohanpur, 741246 India.}
\affiliation {$^2$International Centre for Theoretical Sciences, Tata Institute of Fundamental Research, 
Bengaluru 560089, India.}

\begin{abstract}
 We generate point configurations (PCs) by thresholding the local energy of the Ashkin-Teller model in two dimensions (2D) and study the percolation transition at different values of $\lambda$ along the critical Baxter line by varying the threshold that controls the particle density $\rho$. For all values of $\lambda$, the PCs exhibit power-law correlations with a decay exponent $a$ that remains independent of $\rho$ and varies continuously with $\lambda$.  
For $\lambda < 0$, where the PCs are hyperuniform, the percolation critical behavior is identical to that of ordinary percolation. In contrast, for $\lambda > 0$, the configurations exhibit giant number fluctuations, and all critical exponents vary continuously, but form a superuniversality class of percolation transition in 2D.  
\end{abstract}\maketitle

The central limit theorem (CLT) is a cornerstone of probability theory, asserting that the sum of a large number of independent, identically distributed random variables tends to converge to a normal distribution, irrespective of the original distribution of the variables \cite{Feller1971, Kolmogorov1954, Fischer2011}. This universality underpins much of statistical mechanics and physical sciences, providing a robust framework for analyzing fluctuations in diverse systems.  A simple consequence of  CLT is  that the variance  $\sigma^2$ of the sum depends linearly on the number  of  stochastic variables $V$ summed  over, i.e., $\sigma^2 \sim V^q$ with the fluctuation exponent  $q=1.$ In crystalline and amorphous solids, fluid systems, and beyond, the CLT explains the emergence of Gaussian fluctuations in extensive properties. In  equilibrium systems,  the  linearity  of the variance  along with  the  fluctuation-response relation  assures that the generalized  susceptibilities  $\sigma^2/V$ are  extensive properties  of matter independent of  the  size/volume of the system.   

However, deviations from this  normal  Gaussian  paradigm arise in systems with correlated or constrained dynamics, where   the variance is  either sub-linear ($q<1$)  or super-linear($q>1$) in  number of components $N;$ accordingly the corresponding susceptibilities   vanishes  or diverges.  The first case is  referred to as {\em hyperuniformity} where long-wavelength density fluctuations are suppressed compared to typical disordered systems, 
and the later   one   are  referred to as systems   with giant number fluctuations (GNF), where  fluctuations at macroscopic scales can become anomalously large due to strong correlations or driven dynamics\cite{Ramaswamy2010}.

Evidence of hyperuniformity can be found across a diverse array of natural systems. For instance, the spatial arrangement of photoreceptors in the eyes of birds \cite{bird1,bird2,bird3}, fish \cite{fish}, and other vertebrates \cite{vert} illustrates this property. Similarly, hyperuniform patterns emerge in vegetation distribution within ecosystems \cite{eco1,eco2}, as well as in the layout of human settlements \cite{Human}. In the realm of materials science, examples include the structure of amorphous silica \cite{silica}, vortex lattices in superconductors \cite{vortex}, amorphous ice \cite{ice1}, and binary mixtures of charged colloids \cite{colloid1,colloid2}, leaf vein networks \cite{Liu2024}. Additionally, hyperuniformity is observed in theoretical models, such as those describing two-phase coexistence \cite{2phase1,2phase2,2phase3,2phase4}, self-organized critical states in sandpile models \cite{soc1,eco2}, and other critical absorbing states \cite{Basu,Hexner,CLG_RO,Peter,Lebowitz2024}.
The giant number fluctuations, where the  fluctuation exponent $q>1,$  are also  not uncommon.  It has been observed in active matter systems  \cite{Ramaswamy2003, Narayan2007, Chate2008, Ramaswamy2010, Rajesh2012, Shradha2014, Toner2019, Kuroda2023}, biological systems \cite{Carr2010}, and  some other  critical phenomena.

\begin{figure}[t]
\centering
\includegraphics[width=8.5cm]{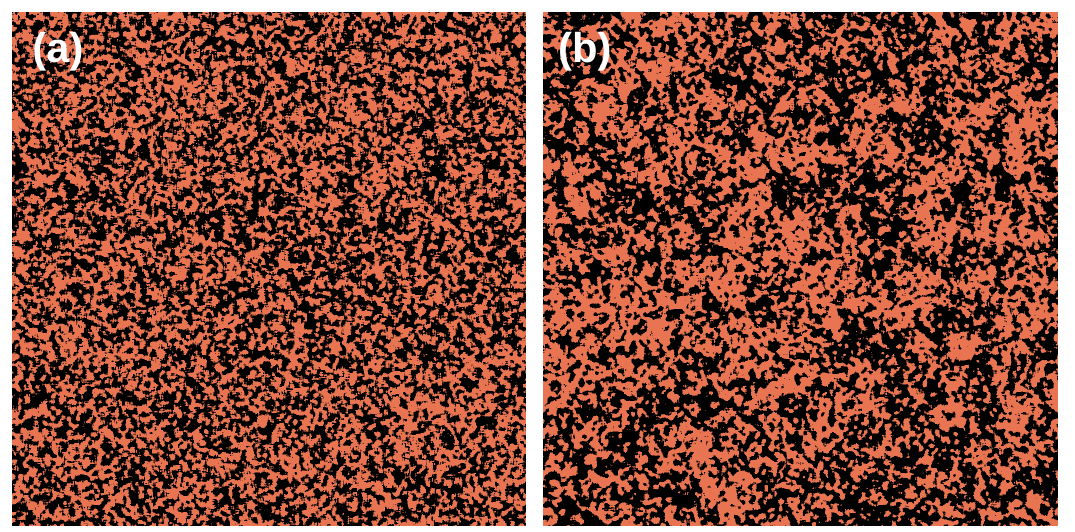}
\caption{Point configurations generated
from  critical Ashkin Teller model by assigning, $1$ (bright) to sites whose energy (coarse grained) exceeds a pre-determined threshold $\eps^*$, else $0$ (dark) on a $1024 \times 1024$ square lattice. Configuration with  (a) $\lambda=-0.2$ (hyperuniform) is compared with (b) $\lambda=0.2$ (giant number fluctuation).  $\eps^*$  is chosen to se particle density is $\rho = 0.5$ for both.}
\label{fig:point_config}
\end{figure}

Despite these widespread occurrences, a comprehensive theoretical framework explaining the stability of hyperuniform structures remains elusive. 
%However, some progress has been made in specific contexts \cite{theory1,theory2}.
 Some models have been proposed  to understand the origin of  hyperuniformity \cite{Lebowitz2024, Punyabrata2024} and GNF \cite{Sachdeva2014, Punyabrata2020} separately. Connecting these concepts requires a nuanced understanding of fluctuation dynamics in complex systems. While hyperuniformity suggests a ``quietness"  \cite{Torquato2018} in certain spectral regions, giant fluctuations underscore the ``wildness" that can arise in far-from-equilibrium \cite{Narayan2007} or highly interactive contexts. 
 A generic theory or dynamics  that  can lead  to a steady state  having  these unusual   fluctuations is far form  reach.  In a recent work \cite{Mukherjee2024}
 the authors have proposed  an elegant method to generate point configurations (PCs) on a  square lattice  from the steady states  of Ashkin Teller (AT) \cite{AT_1943, Baxter} model at criticality where  the fluctuation exponent is $q< 3/4;$  for any given $q$  the density  could be varied  by  imposing a cutoff  on the local energy.  Such a method  can  generate  PCs    can give rise to  hyperuniformity  and GNF  for  different values of the  interaction parameter $\lambda$ on the critical Baxter line \cite{Baxter1971}. A typical point  configurations  for $\lambda=-0.2$ (hyperuniform) is compared  with  the same  obtained for  $\lambda=0.2$ (GNF) in Fig. \ref{fig:point_config} - the threshold  energy  is chosen  such that $\rho=\frac12$ in both cases.

In this article we aims to investigate  the role of unusual fluctuations on  percolation properties of  hyperuniform and  GNF materials.  Since  a  broad spectrum of physical systems are known to  be hyperuniform \cite{Torquato2018},    and  several  new materials   with   these  unusual  properties  have been synthesized  recently \cite{matter1,matter2,matter3} for their  technological applications,  our study of  percolation  is expected to  shed light on understanding  their  geometrical and topological properties,  and  transport
phenomena.

First we discuss how to  generate  point configurations  of any density   having unusual fluctuations  from the critical steady  state of the  
Ashkin Teller (AT) model \cite{AT_1943, Wu_Lin, Kadanoff_1977, Baxter}.
AT model   defined  on a  $L\times L$ square lattice with periodic boundary conditions in both directions.  The sites of the lattice   ${\bf i} \equiv (x,y)$ where $x,y=1,2,\dots, L$   carries two different Ising spins $\sigma_{\bf i}=\pm$ and $\tau_{\bf i}=\pm$  which  interact as following  the Hamiltonian,
\begin{equation} \label{eq:AT_H}
 H = -J\sum_{\langle {\bf ij}\rangle} \sigma_{\bf i} \sigma_{\bf j} - J \sum_{\langle {\bf ij}\rangle}\tau_{\bf i} \tau_{\bf j} - \lambda \sum_{\langle {\bf ij}\rangle}\sigma_{\bf i} \sigma_{\bf j} \tau_{\bf i} \tau_{\bf j}. 
\end{equation}
Here $\langle {\bf ij}\rangle$ denotes a pair of nearest-neighbor sites, $J$ is  the strength of  intra-spin interactions and $\lambda$ represents interactions among  $\sigma$ and $\tau$ spins.  The model undergoes a continuous  phase transition   from a unpolarized paramagnet  to a polarized ferromagnet along  a critical line in $\lambda$-$J$ plane, formally known as the   Baxter line.
The equation of Baxter line,
\be
\sinh(2\beta J) = e^{-2 \beta \lambda}
\label{eq:BL}
\ee
is known exactly from the exact mapping \cite{Fan1970, Kadanoff1971} of AT model and eight vertex (8V) model \cite{Baxter1972}. 

This mapping also provides exact critical exponents \cite{Wu1974, Domany1979} 
%of both magnetic and electric transitions, 
which vary continuously with $\lambda,$ as one moves along the Baxter line; the reason owes to the  marginality  of coupling parameter $\lambda$ \cite{Kadanoff2000}.  Note  that, for $\beta=1$   the critical line  ends  at  $\lambda =\ln(3)/4,$ where  $\lambda=J$ and a new symmetry, $\mathbf Z_4$  appears there. Unless  otherwise specified, in this article we consider $\beta=1$ and remain on the Baxter line   \eqref{eq:BL} to generate the PCs as follows.

\begin{figure}[t]
\centering
\includegraphics[width=4.25cm]{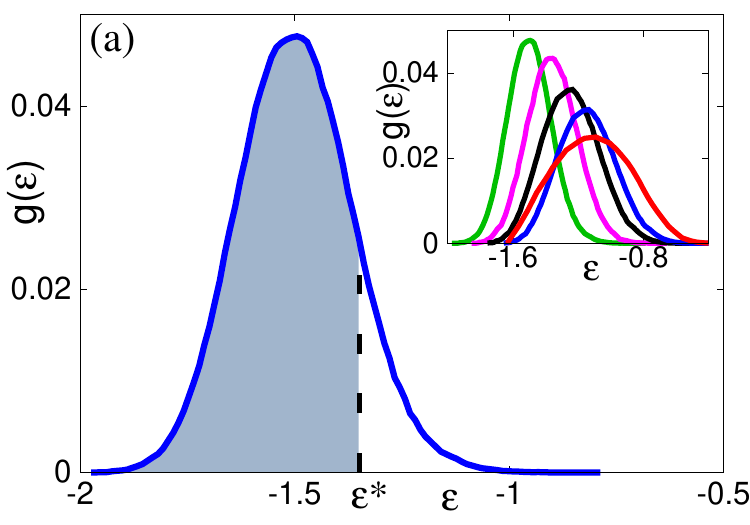}
\includegraphics[width=4.25cm]{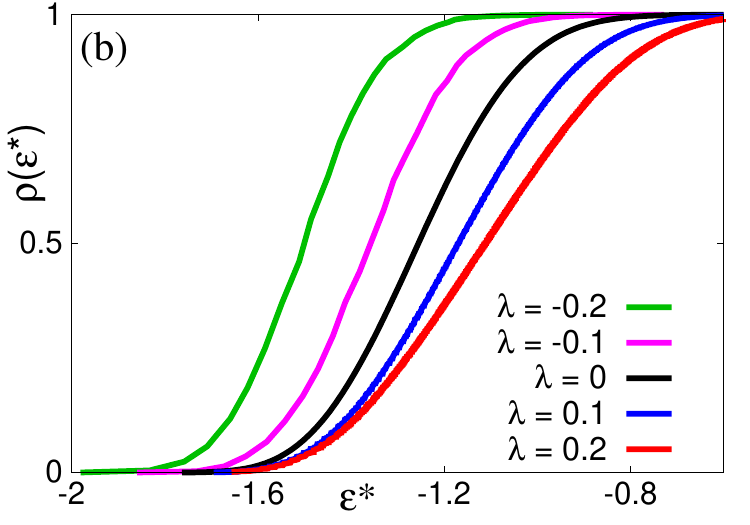}
%\vspace*{-.2 cm}
\caption{(a) Schematic representation  of particle density $\rho(\eps^*),$ as in Eq. \eqref{eq:rhoeps}. The shaded  represents   particle density. Inset shows the distribution  of coarse-grained local energy $\varepsilon$ defined in Eq. \eqref{eq:vareps} with  $l=8.$ Different curves from top to bottom corresponds to  $\lambda = -0.2, -0.1, 0, 0.1,0.2$.
(b) $\rho(\varepsilon^*)$ as a function of the  threshold $\varepsilon^*$ for $\lambda = -0.2, -0.1, 0, 0.1,0.2$.
}
\label{fig:P(e)}
\end{figure}

First we  perform Monte Carlo simulation of the AT model at, 
\be \label{eq:TcJclam}
\beta =1, \lambda_c=\lambda, J_c = \frac12 \sinh^{-1}(e^{-2\lambda}).
\ee
and in  steady state, define a coarse-grained energy at each site {\bf i},
 \be
\varepsilon_{\bf i}\equiv\varepsilon_{x,y}  = \frac{1}{l^2}  \sum_{m,n=0}^{l-1}   H_{x+m,y+n}.
\label{eq:vareps}
 \ee
 Here $(m,n)$ are  positive integers  and $l$  is  the  coarse-graining length-scale  and   $H_{x,y}$  is  the  local energy at each site, 
\bea 
%H_{\bf i} &\equiv& 
H_{x,y} &=& -J s_{x,y} \left( s_{x,y+1} + s_{x+1,y}\right)  -J \tau_{x,y} \left( \tau_{x,y+1} + \tau_{x+1,y}\right)\cr  &-&\lambda \left(  s_{x,y}\tau_{x,y}  (s_{x,y+1} \tau_{x,y+1} +s_{x+1,y}\tau_{x+1,y} \right).
\label{eq:Hxy}
\eea
An occupation variable  $n_{\bf  i}$ is then introduced at  very  site 
\be 
n_{\bf  i} = \theta(\eps^* -\eps_{\bf  i})
\label{eq:n_i}
\ee
where  $\theta(x)$ is  the  Heaviside step function that  takes   the value $0$ when $x$ is $-$ve, and $1$ otherwise and $\eps^*$ is a  energy-threshold  that  controls the density of the  occupied sites, 

\be 
\rho(\eps^*) =\frac{N}{L^2}=
\int\limits_{-\infty}^{\eps^*} g(\eps) d\eps ~~ N=\sum_{\bf i} n_i.
%= \int_{\eps^*}^\infty g(\eps) d\eps.
\label{eq:rhoeps}
\ee
Here  $g(\eps)$ the  probability density function of energy distribution,  shown   in  Fig. \ref{fig:P(e)}(a) for different $\lambda.$  A plot of  density $\rho(\eps^*)$   is  shwon in Fig. \ref{fig:P(e)}(b). Clearly   $\rho(\eps^*),$  being a cumulative density function  increases monotonically with  in crease of the  threshold $\eps^*$ and it can  have a linear form in  the neighborhood  of  $\eps_c$  as  $\rho(\eps^*)  - \rho(\eps_c)\sim  (\eps^*-\eps_c).$ This ensures that  the percolation  critical  exponents  calculated in  terms  $\Delta = \eps^*- \eps_c$  is  same as that of $\rho- \rho_c.$

Note that  $n_i=1,0$  (occupied or empty) is a stochastic variable, and one would expect their sum, $N=  \sum_{\bf i} n_i$  to  have a normal Gaussian distribution, following CLT.  However,   Ref. \cite{Mukherjee2024}  demonstrated that the CLT breaks down in this case, as the number fluctuations in an $l\times l$  system  scales as 
\be 
\sigma^2_l = \langle N_l ^2\rangle  - \langle N_l\rangle^2 \sim l^{2/\nu_{AT}}.
\ee
Since  $V= l^d,$ the fluctuation exponent  is then,  $q= \frac2{d\nu_{AT}}.$ It was also shown that  the exponent does not depend on  the  energy cutoff $\eps^*,$   or equivalently  the  average density $\rho(\eps^*)$ \cite{Mukherjee2024}. From  the exact solution of AT model \cite{Baxter1972, Wu1974, Domany1979} the correlation length exponent $\nu_{AT}$ is known exactly, which results in  a continuous variation   of the   fluctuation exponent  $q$  along the Baxter line,   
%\be q= \frac{2}{d\nu_{AT}}=\frac{2}{d}\frac{4\mu-3\pi}{2 (\mu-\pi)};\mu = \cos^{-1} \left(  e^{2\lambda} \sinh(2\lambda)    \right).  \label{eq:q}\ee
\be q= \frac{2}{d\nu_{AT}}=\frac{2}{d}\frac{4\mu-3\pi}{2 (\mu-\pi)};~~ \mu = \cos^{-1} \left(  e^{2\lambda} \sinh(2\lambda)    \right). \label{eq:q} \ee
Note that for $\lambda=0$   we have  $q=1$  (setting $d=2$) and thus  naturally  the   PCs are hyperuniform ($q<1$) when $\lambda<0$   and they exhibit giant number fluctuations ($q>1$) for any $\lambda >0$.

\begin{figure}[t]
\centering
\includegraphics[width=4.27cm]{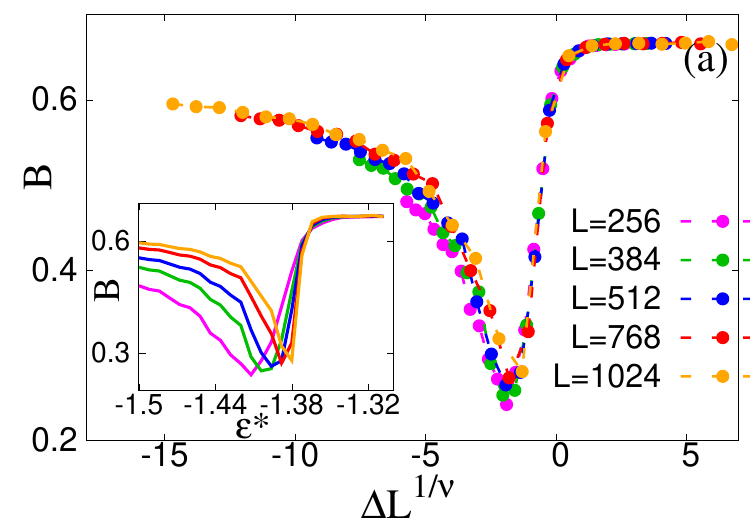}
\includegraphics[width=4.27cm]{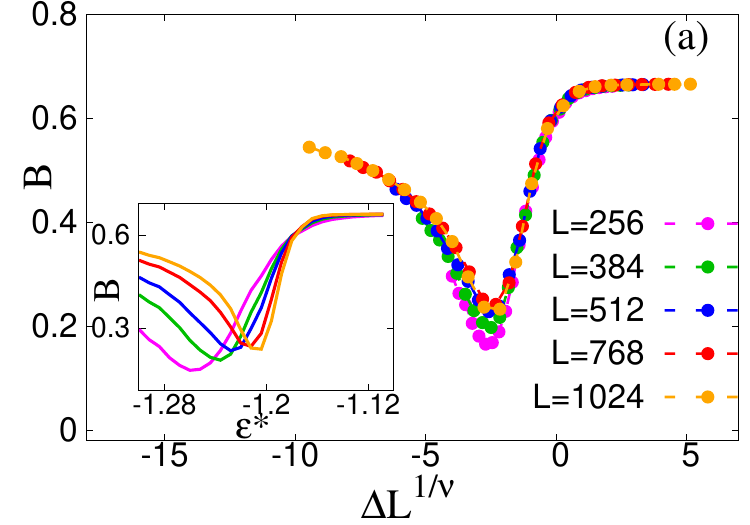}
\includegraphics[width=4.27cm]{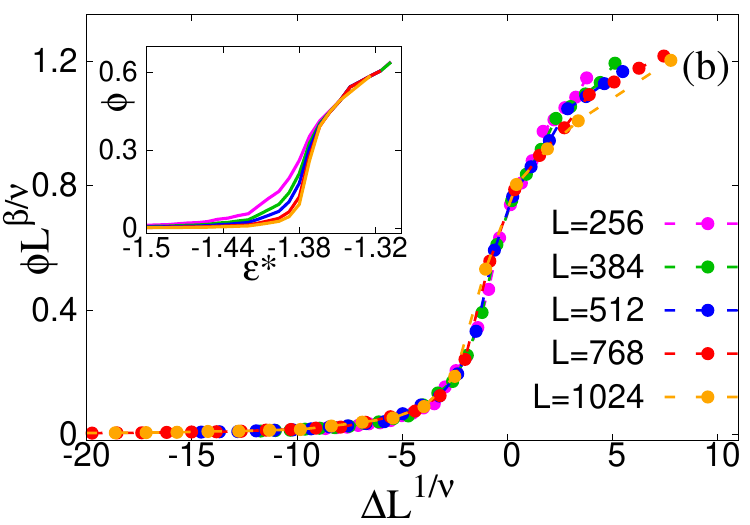}
\includegraphics[width=4.27cm]{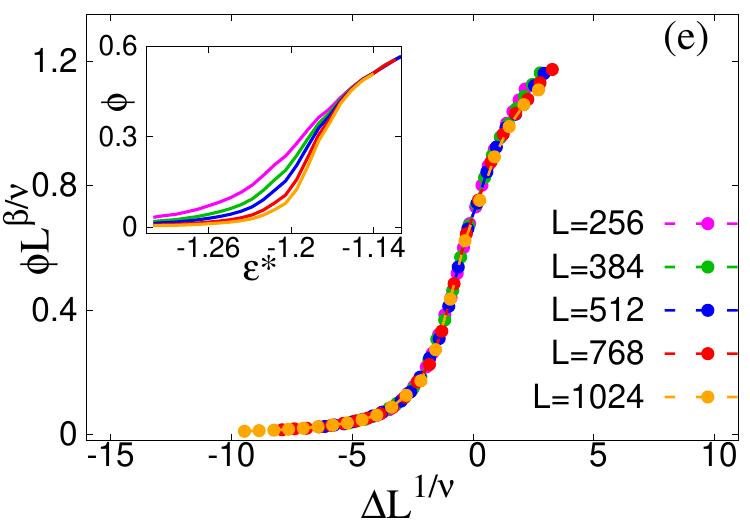}
\includegraphics[width=4.27cm]{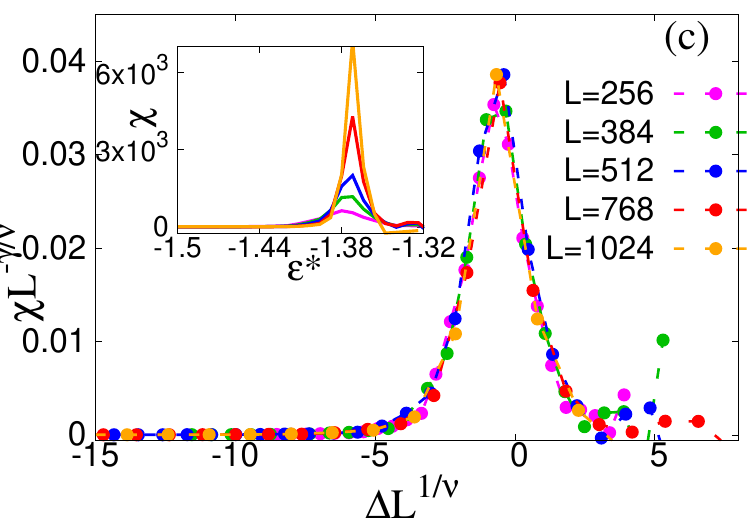}
\includegraphics[width=4.27cm]{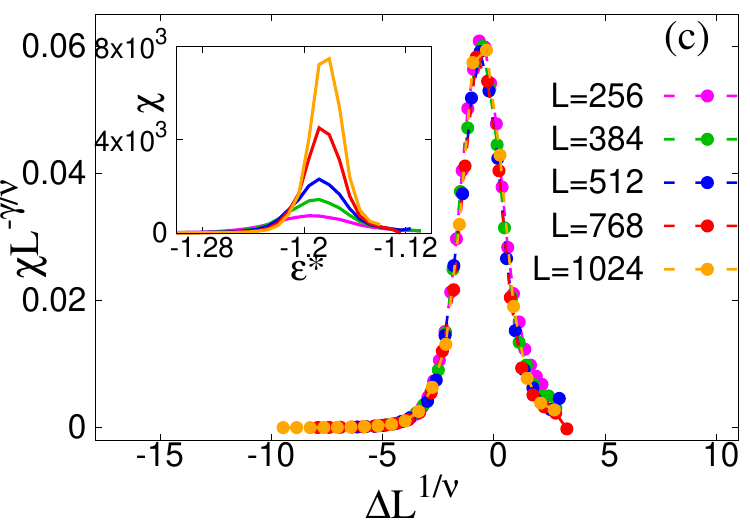}
%\vspace*{-.2 cm}
\caption{ (a) - (c) Data collapse of Binder cumulant $B$, $\phi L^{\beta/\nu}$ and $\chi L^{-\gamma/\nu}$ respectively as a function of $\Delta L^{1/\nu}$, across system sizes $L=256,384,512,768,1024$  to a unique scaling function observed for $\lambda = -0.1$. At the critical threshold $\epsilon^*_c=-1.3664(2),$ the
best collapse is obtained for $ \nu=1.603, \beta = 0.181, \gamma=2.715.$ The uncollapsed  data are shown in the inset. Data are averaged over $10^6$ or more samples in a steady state. (d) - (f) Similar data collapse observed for $\lambda = 0.1$. At the critical threshold $\epsilon^*_c=-1.1757(4),$ the
best collapse is obtained for $ \nu=1.552, \beta = 0.184, \gamma=2.735.$ The uncollapsed plots are shown in the inset. Data are averaged over $10^6$ or more samples in a steady state.}
\label{fig:collapse_lm-0.1_0.1}
\end{figure}
 \begin{figure}[t]
\centering
\includegraphics[width=4.27cm]{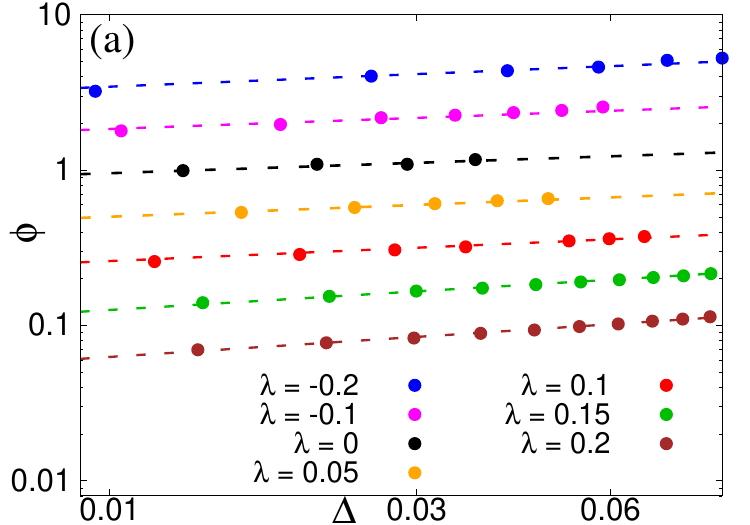}
\includegraphics[width=4.27cm]{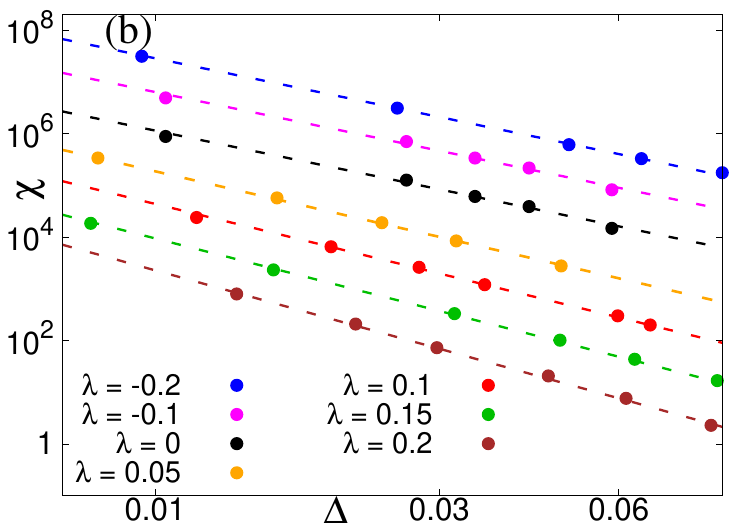}
\includegraphics[width=4.27cm]{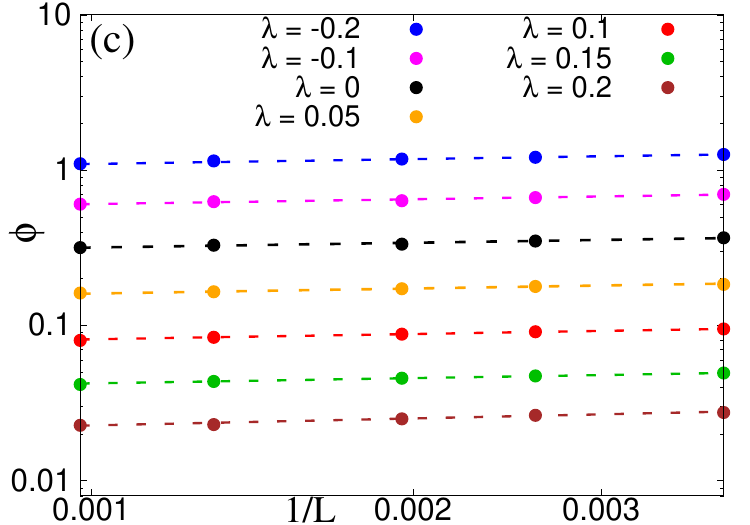}
\includegraphics[width=4.27cm]{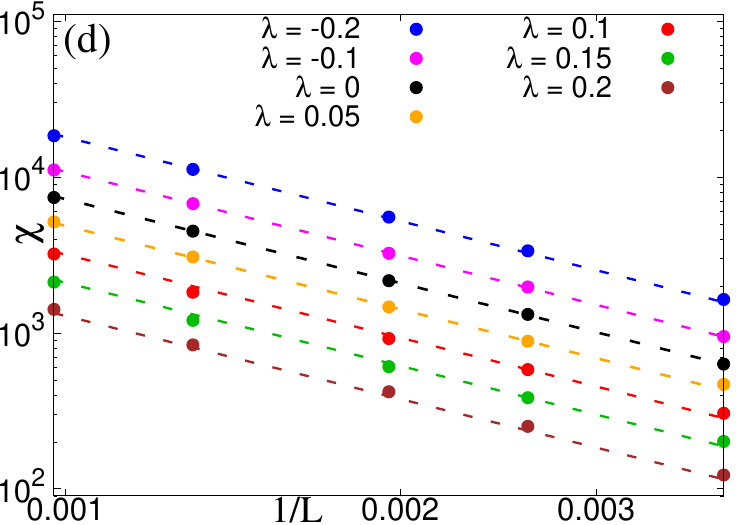}
%\vspace*{-.2 cm}
\caption{Critical exponents from log-scale plots: (a) - (b) $\beta, \gamma$ from the scaling of the order parameter $\phi$ and its second cumulant $\chi$ with $\Delta$ at the critical threshold $\epsilon^*=\epsilon_c$ for different $\lambda$ values and $L=1024.$ (c) - (d) Exponents $\beta/\nu, \gamma/\nu$ from the scaling of $\phi$ and $\chi$ with $1/L$ at the critical point $\epsilon_c$ for different values of $\lambda.$ Points are from numerical simulation. Dashed lines with slopes are the exponent values, drawn for comparison. In all cases, data is averaged over $10^6$ or more samples, and the $y$-axis is
scaled by factors $\{\frac{1}{4},\frac{1}{3},\frac{1}{2},1,2,3,4\}$ (from bottom to top) for better visibility.  The critical exponents obtained here are listed in Table \ref{tab:table1}. Error bars are the same size or smaller than the symbols used.
}
\label{fig:exponents}
\end{figure}
 %%%%%%%%%%%%%%%%%%%%%%%%%%%%%%%%%%
\begin{table*}[t]
\caption{\label{tab:table1}
Estimated values of the critical point and critical exponents along AT critical line for different $\lambda$
}
\begin{ruledtabular}
\begin{tabular}{lcccccccccr}
\textrm{$\lambda$}&
\textrm{$\nu_{AT}$}&
\textrm{$a$}&
\textrm{$\eps_c$}&
\textrm{$\rho_c$}&
\textrm{$\beta$}&
\textrm{$\nu$}&
\textrm{$\gamma$}&
\textrm{$\frac{\beta}{\nu}$}&
\textrm{$\frac{\gamma}{\nu}$}\\

\colrule
-0.2 &1.275&2.432  &-1.4975(2) &0.4997(1) & 0.138(6) & 1.332(9) & 2.384(8) & 0.103(9) & 1.794(2)  \\
-0.1 &1.134& 2.236 &  -1.3664(2) & 0.4986(4) & 0.138(9) & 1.333(0) & 2.387(9) & 0.103(7) & 1.794(9) \\
0 &1&	2 & -1.2525(2) & 0.4982(9) & 0.138(8) & 1.333(2) & 2.388(4) & 0.104(1) & 1.795(6) \\
0.05 &0.938& 1.868 &  -1.2060(2) & 0.4981(0) & 0.157(7) & 1.434(1) & 2.571(2) & 0.109(9) & 1.781(1) \\
0.1 &0.870& 1.727 &  -1.1757(4) & 0.4874(3) & 0.184(9) & 1.552(7) & 2.735(1) & 0.112(6) & 1.764(2) \\
0.15 &0.825& 1.575 &  -1.1538(3) & 0.4727(7) & 0.229(2) & 1.709(3) & 2.983(3) & 0.145(9) & 1.743(6) \\
0.2 &0.772& 1.409 &  -1.1437(1) & 0.4671(1) & 0.271(4) & 1.890(7) & 3.182(4) & 0.152(0) & 1.685(7)\\
\end{tabular}
\end{ruledtabular}
\end{table*}
%%%%%%%%%%%%%%%%%%%%%%%%%%%%%%%%%%
 Now that the PCs  with hyperuniform  and GNF  can be  produced with  any  desired $q$ (by varying   $\lambda$ along  the Baxter line),  and  density $\rho$  (by  varying $\eps^*$), we  proceed to  study  site-percolation  phenomena on these PCs.
 In site percolation, two neighboring sites of a lattice are considered part of the same cluster if they are both occupied. If  there are $K$-clusters in a configuration labeled by $k=1,2\ldots,K,$ each having  $s_k$ number of particles then $\sum_k s_k =\sum_{\bf i} n_i=N =\rho L^2,$  where $\rho$  is the  particle density.  Let us denote  size of the largest cluster  in a configuration  as  $s_{max}= {\rm max} (\{s_k\}).$   It is a  standard practice in site-percolation transition to consider   the  mean density  of the  largest  cluster, $s_{{\rm max}}/L^2,$ as the order parameter  of the system  because  $s_{max}$ become macro-size   at the critical density  $\rho_c$.  In AT model     the density $\rho$  is a  function  of  $\eps^*,$ as  described in  Eq. \eqref{eq:rhoeps}, and percolation occurs when  $\eps^*$  reach a critical value $\eps_c.$

In the following we study  scaling properties   of the   
order parameter $\phi,$ its  variance   $\chi$  and the  fourth order cumulant, formally known as the  Binder cumulant $B.$   In the near critical zone  they scale as \cite{Stanley1972, Cardy},
\bea
\phi &=&  \frac1{L^2}\langle s_{max} \rangle\sim |\Delta|^\beta ;
\chi =\frac{1}{L^2} (\langle s_{max} \rangle -\langle s_{max} \rangle) \sim   |\Delta|^\gamma\cr 
B&=& 1 -\frac13\langle s_{max}^4 \rangle/\langle s_{max}^2 \rangle^2.
\label{eq:Scalling}
\eea
where $\Delta= \eps^* - \eps_c.$ These two exponents $\beta,\gamma$  along with    $\nu,$  associated with the scaling of the correlation length $\xi\sim |\Delta|^{-\nu},$  are  characteristic features of the percolation  universality class. In 2D, the  percolation  exponents \cite{Stauffer2018} are $\nu= \frac43,$ $\beta = \frac{5}{36},$  and $ \gamma= \frac{43}{18} .$ To calculate the critical exponents   of 
 the percolation transition  of PCs  generated  from critical AT model,  we   employ   finite size scaling (FSS) analysis \cite{Stanley1972, Cardy}.
 \begin{eqnarray}
 \label{eq:fss}  
 B = f_{b}(\Delta L^{\frac1\nu});~~
 \phi = L^{-\frac{\beta}{\nu}}f_\phi(\Delta L^{\frac1\nu});~~
 \chi = L^{\frac\gamma\nu}f_\chi(\Delta L^{\frac1\nu})
 \nonumber
\end{eqnarray}
using  $\eps^*$ as  the   control parameter and  $\Delta = \eps^* - \eps_c;$  $\eps_c$ is the critical  energy threshold  which corresponds  to a  definite average density $\rho(\eps_c).$

From  the Monte Carlo simulation  of  AT model, for a fixed value of $\lambda$ on the Baxter line,  we generate PCs  by  thresholding   local energy  at  $\eps^*$   using Eq.  \eqref{eq:n_i}.  Number of clusters $K,$ their sizes $\{s_i\}$ and  the maximum cluster $s_{max}$  in  the  PCs  are  calculated  using Newman-Ziff algorithm\cite{Newman2000}.  From  the  moments of the   $s_{max}$  we obtain  $\phi$, $\chi$, and $B$  for different  $\eps^*$ using  Eq. \eqref{eq:Scalling} and  then repeat the  process for different $L.$   Since, at the critical point,  the Binder cumulants  does not depend on system size  we identify $\eps_c$ as the crossing point of $B$ vs. $\eps^*$  curves for different $L,$ and note the average density $\rho_c$ there. The plots of  $B,$  $\phi L^{\beta/\nu}$ and  $\chi L^{-\gamma/\nu}$ as a function of $\Delta L^{1/\nu}$   are then adjusted  by tuning $\frac1\nu,\beta/\nu, \gamma/\nu$  respectively  to achieve the best data collapse, and get the estimate of the  exponents.   These  exponents, obtained for different $\lambda$ are listed in Table \ref{tab:table1}. The data collapse, for $\lambda=-0.1$ (hyperuniform)  and  $\lambda=0.1$  (GNF) are demonstrated in Fig. \ref{fig:collapse_lm-0.1_0.1}. Similar figures for other $\lambda$ values are  shown in the Supplemental  Material \cite{SM}.\\

We further check  that  the estimated   exponents are  consistent with  $\beta$ and $\gamma$    obtained  directly   from the scaling: $ \phi \sim \Delta^{\beta} ; \chi \sim \Delta^{-\gamma}$ for a   large $L.$   The log-scale plot  of  $\phi, \chi$  as  a function of $\Delta$  for different $\lambda$   are  shown in Fig. \ref{fig:exponents}(a),(b) respectively  for $L=1024.$  In a similar way exponents $\frac\beta\nu, \frac\gamma\nu$  can be obtained directly from  using   the relations  at the critical point $\eps^*=\eps_c$: $\phi\sim L^{\frac{\beta}{\nu}},   \chi  \sim L^{-\frac{\gamma}{\nu}}.$  The  log-scale plots of  $\phi,\chi$ as  a  function of $L$ are shown in    \ref{fig:exponents}(c),(d)
for different $\lambda.$   In all these  figures, the   guiding lines along the data  corresponds to the  exponents  estimated from finite size scaling - they match  quite well.   Our  final estimate of  the critical   energy threshold, corresponding density and  the  critical    exponents are listed in  Table \ref{tab:table1}. 

\begin{figure}[t]
\centering
\includegraphics[width=4.26cm]{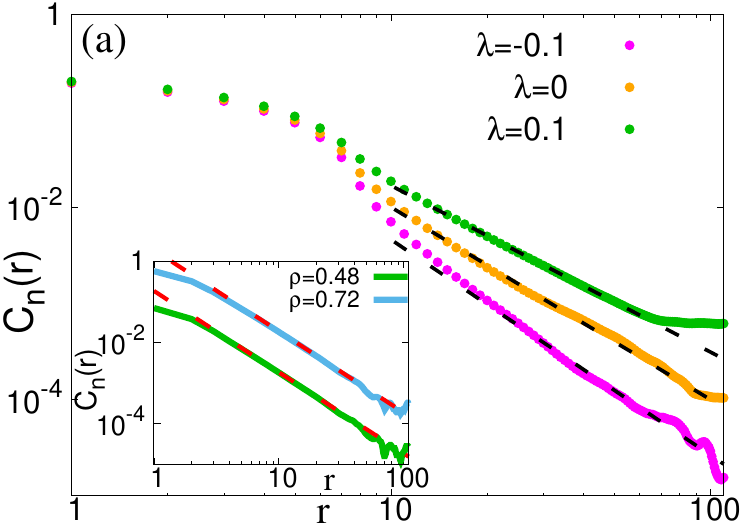}
\includegraphics[width=4.26cm]{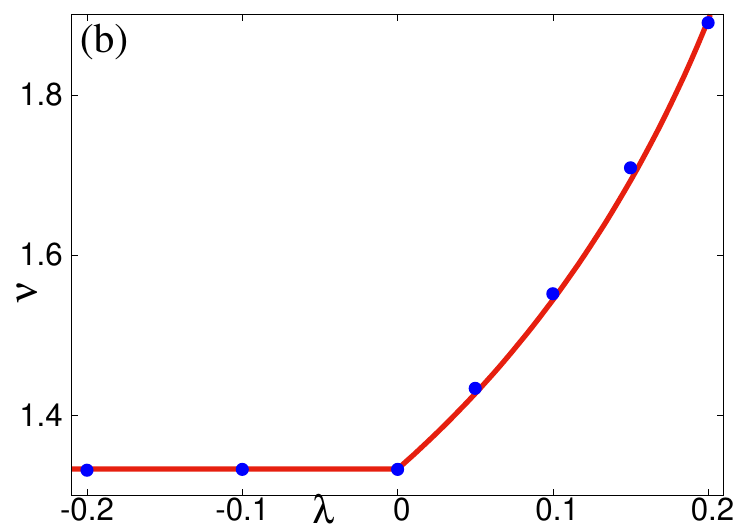}
%\vspace*{-.2 cm}
\caption{(a) Correlation function $C_n(r)$ of the point configurations for different values of $\lambda = -0.1,0,0.1.$ Points are from the numerical simulation, and the dashed lines, depicting the value of the correlation exponent $a$, are drawn for comparison. For $\lambda=-0.1,0,0.1$ the value of the exponents are $2.24, 2$ and $1.73$ respectively. In all cases the simulation has been performed for $L=1024$ and the data is averaged over $10^6$ or more samples. Inset shows the correlation function $C_n(r)$ for $\lambda=0$ with two different density values $\rho=0.48, 0.72.$ Here $L=512,$ the exponent $a$ remains same for both the cases. The $y$ axis is scaled by factors $\{1,2\}$ (from bottom to top) for better visibility. In  (b) Critical exponent $\nu$ obtained from Monte Carlo simulations (symbols) are compared with their exact theoretical
predictions given in Eq. \eqref{eq:nu_a} (solid lines).}
\label{fig:cr}
\end{figure}

It is evident from  Table  \ref{tab:table1} that the critical exponents  for $\lambda<0$  (hyper-uniform PCs) are, within the error limits,  same as   that  of  the ordinary  percolation transition of randomly placed particles on a lattice.  On the  other hand the critical exponents  for    PC having  GNF  ($\lambda>0$)  vary continuously. %This can be explained as follows.
To understand the  nature of  of continuous exponents variation  we  exploit the connection of the fluctuation exponent $q$  with the  correlation exponent $a$ of the PC defined by, 
\be
C_{n}(r) = \langle n_{\bf i} n_{\bf i+r}\rangle -  \langle n_{\bf i} \rangle\langle   n_{\bf i+r}\rangle \sim \frac{1}{r^a},
\label{eq:cn}
\ee
where $\xi$ is the correlation length and $r= |\bf r|.$ Since  the   occupation variables $\{ n_{\bf i}\}$    are generated    from  thresholding  the  coarse-grained local variables  $\{ \eps_{\bf i}\},$   exponent $a$ must be  related to   the exponent associated  with the energy correlation function of AT model, 
\be
C_{AT}(r) = \langle \eps_{\bf i}  \eps_{\bf i+r}\rangle -  \langle  \eps_{\bf i} \rangle\langle    \eps_{\bf i+r}\rangle \sim \frac{1}{r^{d-2+\tilde \eta_{AT}}}, 
\label{eq:en}
\ee
where $\tilde \eta_{AT}$   is different from the  order-parameter correlation exponent $\eta_{AT}.$ 
Since the PCs are generated  from  local energy  we expect  the particle correlation function  to follow   the same power-law decay  $C_n(r) \sim r^{-a},$ with  $a= d- 2 + \tilde \eta_{AT}.$ From the hyper-scaling relations $ \tilde \eta_{AT} = 2-\frac{\alpha_{AT}}{\nu_{AT}},$ and  the exact results   of AT model \cite{Wu1974, Domany1979}  
 $\alpha_{AT} = 2(1-\nu_{AT})$  we find that in   two dimension ($d=2$), 
\be 
a = 2-\frac{\alpha_{AT}}{\nu_{AT}}=  4- \frac 2{\nu_{AT}} = \frac{7\pi}{\mu-\pi}
\label{eq:a}
\ee
where $\mu = \cos^{-1}  e^{2\lambda} \sinh(2\lambda).$  We verified this from  Monte Carlo simulations of  AT  model that generates correlated PCs. Log-scale plot of $C_n(r)$  vs. $r$ are shown in  Fig. \ref{fig:cr}(a)  for  $\lambda=-0.1, 0, 0.1$. Straight lines of slope $2.24, 2, 1.73$  (calculated  from Eq. \eqref{eq:a})  are drawn along the data  for comparison; they  match very well. 

Percolation of  correlated  PCs  have been  studied  extensively \cite{Weinrib1984, Zierenberg2017, Prakash1992, Zierenberg2017, Schrenk2013}. A correlated point configuration  can be treated as a perturbation to the  ordinary un-correlated  point configurations,  with the deviation acting as quenched disorder. The  primary question is  then, whether quenched disorder  is a  relevant  perturbations  to a  continuous phase transition.   A general rule, commonly known as  Harris criterion \cite{Harris}   states that the  disorder is a relevant perturbation when $d\nu_0<2,$   $\nu_0$   being   the correlation length exponent of a clean (disorder-free) system in $d$ dimensions.  The criteria is further  extended for long-range  systems in a seminal work  by  Weinrib and  Halperin (WH) \cite{Weinrib1983}: disorder is  irrelevant  when  $a<d$  and $a\nu>2$,  or when  $a>d$  and $d\nu>2.$  They also show that the  relevant disorder  brings in continuous variation of $\nu$  with correlation exponent $a$ as  $\nu= \frac2a.$  In our problem  we have   two different  correlation  length exponents, $\nu_{AT}$ and $\nu,$    respectively from  the  critical Baxter line   on which  the PCs are generated  and   from the percolation  transition.
%with continuously varying $\nu_{AT}$  we can consider $\lambda=0$ is the  clean  system, where  AT model has $\nu_{AT}=1$ whereas the PCs generated   there exhibits $\nu=\frac43,$ same as the correlation length exponent of ordinary percolation transition  in 2D.  
If we consider $\nu,$ the  WH criterion  predicts   the disorder   to be  relevant for $a<\frac32,$   which    is  indeed 
 observed  in  percolation transition of  PCs  with power-law  correlation $\sim r^{-a},$  generated  by  other methods \cite{Makse1996}. Our observation  however contradicts  it -we  see relevant  changes in   percolation critical behaviour for $a<2,$   which is consistent   when  we consider  $\nu_{AT}$
 for  WH criterion: then disorder is  irrelevant when  $a>d=2$ and $d\nu_{AT} >2 \equiv  \nu_{AT}>1.$  However,  the  observed value of $\nu$ is  not  simply  $\frac{2}{a}$ rather  it is $\frac{2}{a}\nu_0$   where 
 $\nu_0=\frac43$ is the  correlation length exponent of ordinary percolation transition. We propose, 
\be 
 \nu=\left\{ \begin{matrix}  \frac{2}a \nu_0 &0\le a< \frac2{\nu_{AT}}  &{\rm relevant}\\ \nu_0  &{\rm irrelevant} \end{matrix} \right..
 \label{eq:nu_a}
\ee
In   Fig. \ref{fig:cr}(b)   we plot   $\nu$ obtained from   simulations  (symbols) along with   Eq. \eqref{eq:nu_a}; they match  quite well.

In summary, we study the percolation transition of point configurations (PCs) exhibiting unusual number fluctuations, $\sigma^2 \sim V^q$, where $q < 1$ represents hyperuniformity, and $q > 1$ corresponds to systems with giant number fluctuations. The PCs are generated by applying a threshold $\eps^*$ to the critical energy profile of the Ashkin-Teller model in 2D, where $q = \frac{2}{d \nu_{AT}}$ varies continuously along the critical line as the inter-spin interaction parameter $\lambda$ is varied.  
We find that the site percolation exponents remain unchanged when the PCs become hyperuniform ($q < 1$). In contrast, giant number fluctuations ($q > 1$) alter the exponents, causing them to vary continuously.  

\begin{figure}[t]
\vspace*{.3 cm}
\centering
\includegraphics[width=4.265cm]{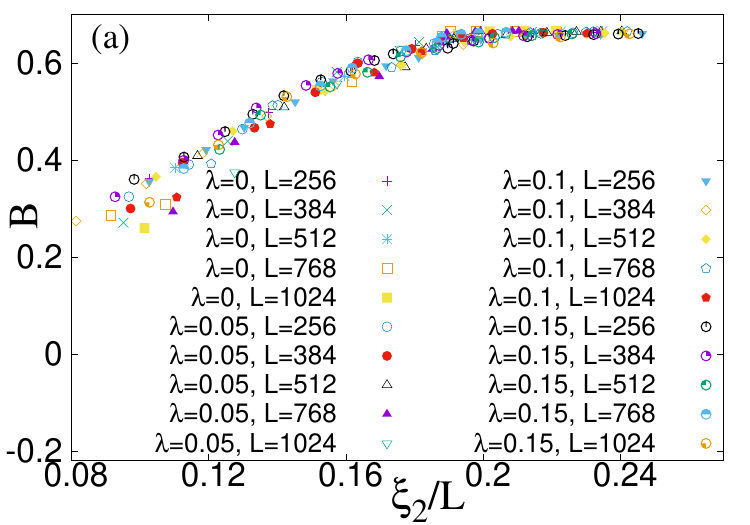}
\includegraphics[width=4.265cm]{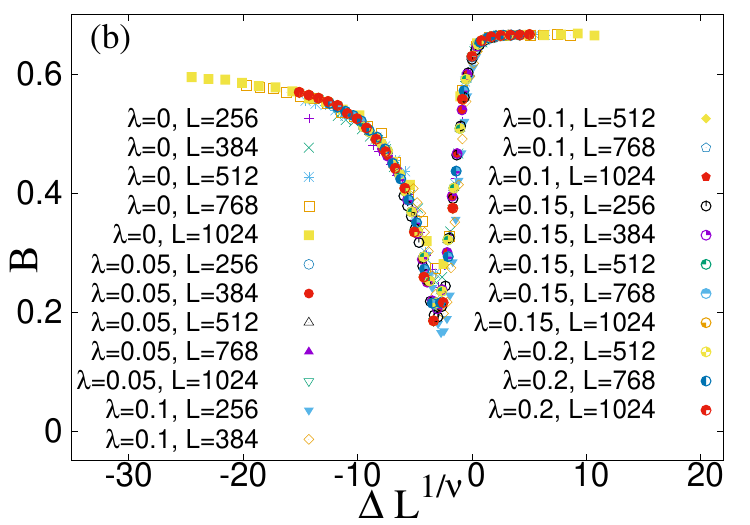}
%\vspace*{-.2 cm}
\caption{(a) Binder cumulant  $B$ vs $\xi_2/L$, (b) $B$ vs $\Delta L^{1/\nu}$. We consider five different $\lambda$ values $\{0, 0.5 0.1,0.15 0.2\}$
and several values of $L$ for each $\lambda$. $\xi_2/L$ is the second-moment correlation length defined in Eq. \eqref{eq:xi2}. In both cases, the data appear to converge into a unique scaling curve. Data are averaged over $10^7$
samples. Error bars are the same size or smaller than the symbols
used.
}
\label{fig:binder_xi}
\end{figure}
Fluctuation exponents are generally related to correlation exponents through scaling relations. In this case, the correlation exponent is given by  $a= 4 - \frac{2}{\nu_{AT}} = 2(2-q),$  
which indicates that disorder becomes relevant when
$q>1$  or equivalently $\nu_{AT}>1.$ This condition is the same as the well-known Harris criterion $d\nu_{AT}>2.$ At the same time, $q>1$  also implies that percolation criticality is altered when $a<2,$ which differs from earlier claims suggesting disorder is relevant for $a<3/2$ \cite{Prakash1992, Weinrib1984}. However, we argue that 
$a$ alone is insufficient to determine whether disorder is relevant; the underlying dynamics that generate correlations also play a crucial role. Specifically, when the underlying dynamics are critical, the same $a$ value can emerge from two different universality classes that differ in other exponents. In such cases, the criterion for assessing the effect of disorder on the system must be reconsidered.

In the case of GNF, the critical percolation exponents vary continuously, raising the question of whether they form a superuniversality class of site percolation. The superuniversality hypothesis suggests that in critical systems with continuously varying critical exponents, certain scaling functions remain identical to those of the parent universality class \cite{Vicari, suh}. One such scaling function is the Binder cumulant $B,$ expressed as a   function of $\xi_2/L$  where  $\xi_2$ represents the second-moment correlation length. 
\begin{equation}
\begin{aligned}
    \xi_2 = \left(\frac{\int_0^\infty r^2C_n(r)\,dr}{\int_0^\infty C_n(r)\,dr}
\right)^{\frac12}
\end{aligned}
\label{eq:xi2}
\end{equation}   

 We  calculate  $\xi_2$   from  the  correlation  function $C_n(r),$  obtained from from Monte Carlo simulations for  different  $\lambda$   and $\eps^*$  near  the critical   Baxter line. The calculations are repeated for different system sizes $L.$   Binder cumulant $B,$   plotted against $\xi_2/L$ as a parametric function of  $\eps^*,$   is shown   in Fig. \ref{fig:binder_xi}(a)  for different $\lambda$ and $L.$  
 The curves match well with each other and also align with the corresponding function obtained for ordinary site percolation.
A similar  plot of  $B$ as a function of $\Delta L^{1/\nu}$ (equivalent to $(\xi/L)^{-1/\nu}$)
also exhibits good collapse. This provides clear evidence that the percolation transition of PCs exhibiting GNF forms a superuniversality class of ordinary percolation.

 We must emphasize that the percolation transition reported here is fundamentally different from the geometric percolation of spins and spin-dipoles studied by Banerjee et al. \cite{Banerjee2025}. While that work addressed whether spins and spin-dipoles - which give rise to simultaneous magnetic and electric transitions along the Baxter line - exhibit magnetic and electric percolation transitions and their universality classes, our focus here is on the coarse-grained local energy. This energy field, while analytic at the critical line, can nevertheless exhibit a percolation transition when thresholding is used to generate point configurations of arbitrary density.

Our findings in the context of percolation studies suggest that hyperuniform disorder is irrelevant, whereas disorder with giant fluctuations is relevant and alters the critical exponents of the clean (disorder-free) system. This insight could be more general and may apply to other critical behaviors as well. Notably, this observation aligns with the well-known Harris criterion \cite{Harris, Weinrib1983} regarding the effect of disorder on a system's critical behavior. Hyperuniform systems appear disordered at small length scales but behave like an ordered crystal as the length scale increases. In the presence of a diverging length scale (such as on the critical Baxter line), the PCs tend to appear ordered at the macroscopic scale. It is, therefore, natural to expect that hyperuniform disorder remains irrelevant to criticality. In continuous percolation and in systems where particles are randomly displaced from their original lattice sites—leading to a presumably hyperuniform configuration—the percolation exponents remain unchanged \cite{Sayantan2019, Mertens2012}.
On the other hand, configurations with GNF ($q>1$) exhibit large-scale fluctuations that dominate the existing statistical fluctuations near criticality. It is therefore unsurprising that disorder characterized by GNF alters the critical behavior of the system.\\

{\bf Author contribution:} SM and IM  contributed equally to this work;  both are  considered co-first authors. \\

{\bf Acknowledgement:} SM gratefully acknowledges financial support through a National Postdoctoral Fellowship from
the Anusandhan National Research Foundation (ANRF), Department of Science and Technology, Government of India, under project File no :  PDF/2023/002952.

\bibliographystyle{apsrev4-2}

\newpage
\setcounter{equation}{0}
\setcounter{figure}{0}

\renewcommand{\theequation}{S\arabic{equation}}
\renewcommand{\thefigure}{S\arabic{figure}}
\renewcommand{\bibnumfmt}[1]{[S#1]}
\renewcommand{\citenumfont}[1]{S#1}

\onecolumngrid
\clearpage
\begin{center}
{\Large {\bf Supplemental Material for ``Percolation of point configurations with  hyperuniform or giant number-fluctuations"}}
\end{center}
\vspace{0.5cm}

\begin{center}
In this supplemental material, we present on estimating the critical point and the critical exponents of site percolation of point configurations in the Ashkin-Teller model for different values of $\lambda$.
\end{center}

\begin{figure}[h]
\centering
\includegraphics[width=5.6cm]{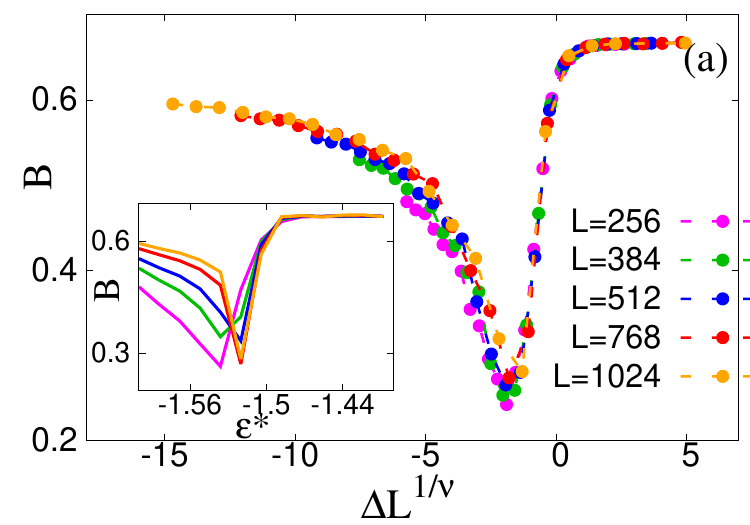}
\includegraphics[width=5.6cm]{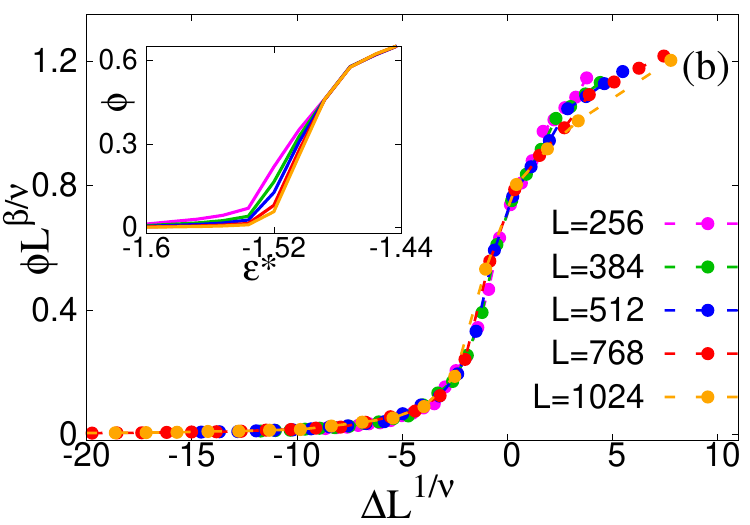}
\includegraphics[width=5.6cm]{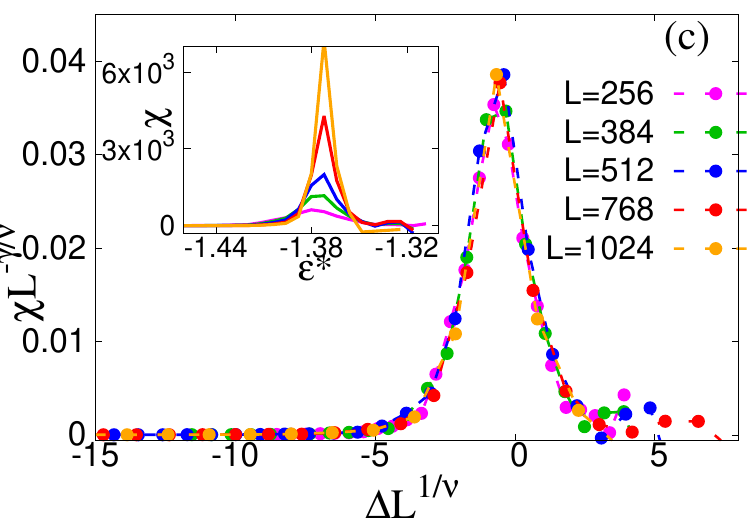}
\vspace*{-.2 cm}
\caption{ Data collapse of (a) Binder cumulant $B$, (b) $\phi L^{\beta/\nu}$ and (c) $\chi L^{-\gamma/\nu}$ as a function of $\Delta L^{1/\nu}$, across system sizes $L=256,384,512,768,1024$  to a unique scaling function observed for $\lambda = -0.2$. At the critical threshold $\epsilon^*_c=-1.4975(2),$ the
best collapse is obtained for $ \nu=1.332, \beta = 0.138, \gamma=2.384.$ The uncollapsed plots are shown in the inset. Data are averaged over $10^6$ or more samples in a steady state.}
\label{fig:collapse_lm-0.2}
\end{figure}

\begin{figure}[h]
\centering
\includegraphics[width=5.6cm]{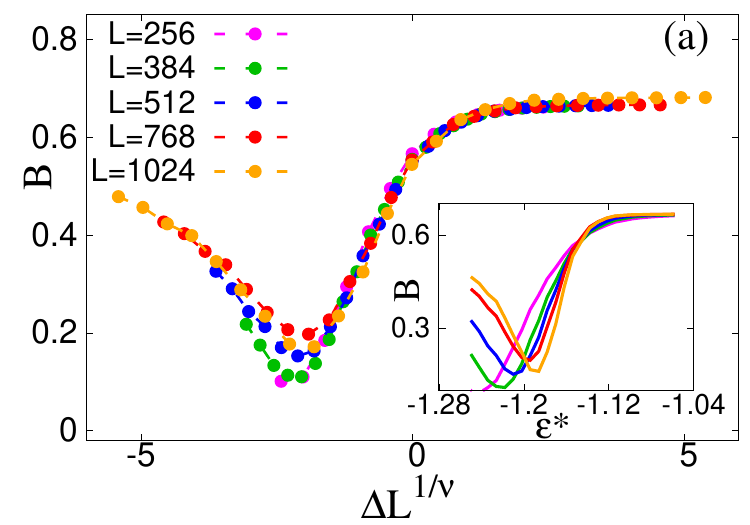}
\includegraphics[width=5.6cm]{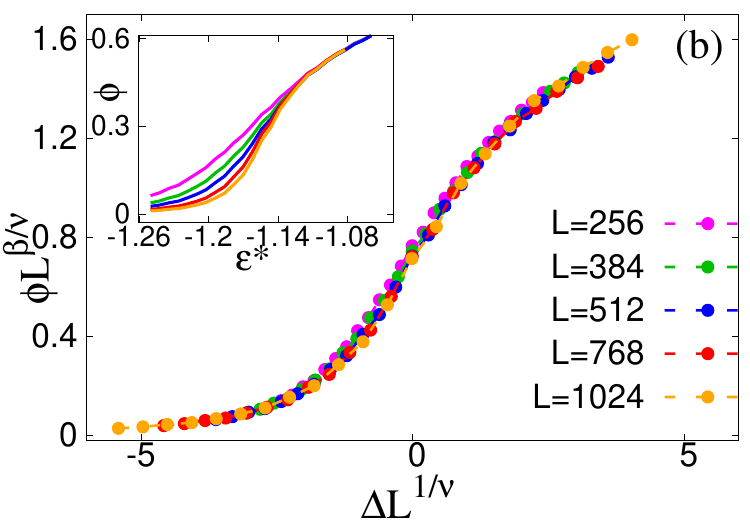}
\includegraphics[width=5.6cm]{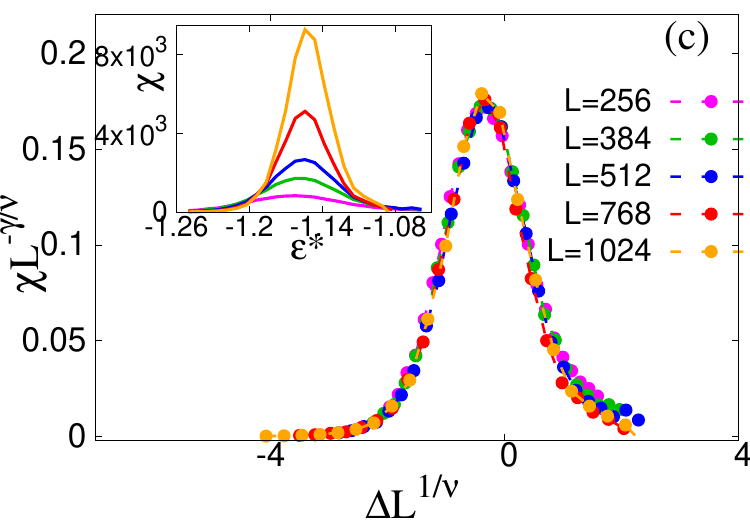}
\vspace*{-.2 cm}
\caption{Data collapse of (a) Binder cumulant $B$, (b) $\phi L^{\beta/\nu}$ and (c) $\chi L^{-\gamma/\nu}$ as a function of $\Delta L^{1/\nu}$, across system sizes $L=256,384,512,768,1024$  to a unique scaling function observed for $\lambda = 0.15$. At the critical threshold $\epsilon^*_c=-1.1538(3),$ the
best collapse is obtained for $ \nu=1.709, \beta = 0.229, \gamma=2.983.$ The uncollapsed plots are shown in the inset. Data are averaged over $10^6$ or more samples in a steady state.
}\label{fig:collapse_lm0.15}
\end{figure}

\begin{figure}[h]
\centering
\includegraphics[width=5.6cm]{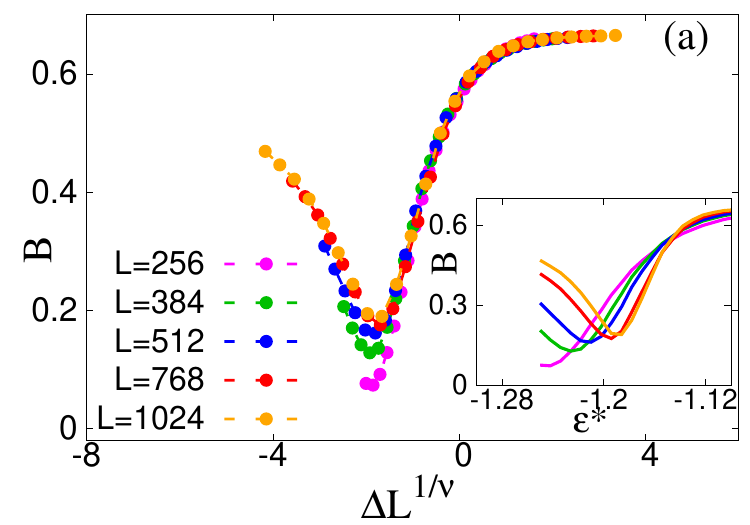}
\includegraphics[width=5.6cm]{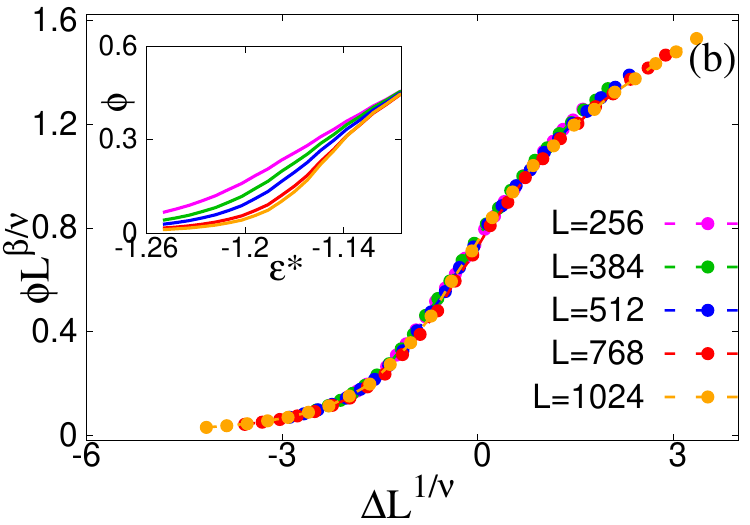}
\includegraphics[width=5.6cm]{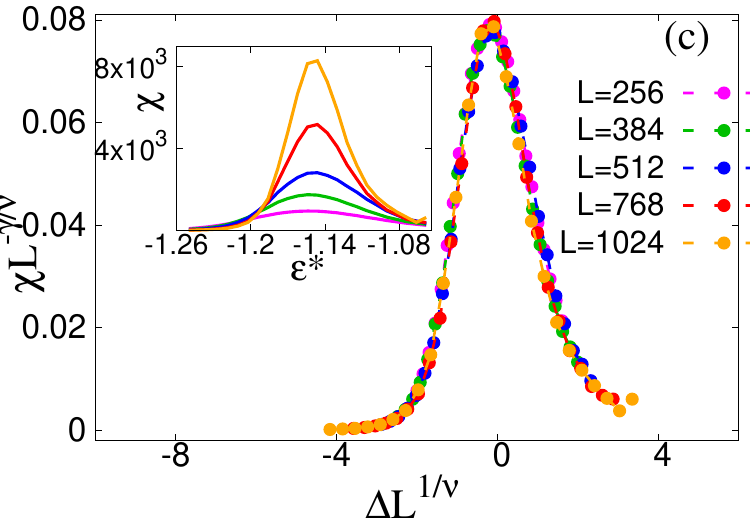}
\vspace*{-.2 cm}
\caption{Data collapse of (a) Binder cumulant $B$, (b) $\phi L^{\beta/\nu}$ and (c) $\chi L^{-\gamma/\nu}$ as a function of $\Delta L^{1/\nu}$, across system sizes $L=256,384,512,768,1024$  to a unique scaling function observed for $\lambda = 0.2$. At the critical threshold $\epsilon^*_c=-1.1437(1),$ the
best collapse is obtained for $ \nu=1.890, \beta = 0.271, \gamma=3.182.$ The raw  data are shown in the inset. Data are averaged over $10^6$ or more samples in a steady state.
}\label{fig:collapse_lm0.2}
\end{figure}

\vspace{0.5cm}
The description of the point configuration (PC) in the Ashkin-Teller model, as obtained from Monte Carlo simulations, is described in Eqs. \eqref{eq:vareps} - \eqref{eq:n_i}.  As the PC with hyperuniformity ($q<1$) and giant number fluctuation ($q>1$) can be generated along the Baxter line for different values of $\lambda$  that governs  the fluctuation exponent $q$;  for any $q,$ any  particle density  $\rho$  can be  obtained by changing 
 the threshold  energy $\epsilon^*.$ We investigate the site percolation phenomenon on these PCs. In site percolation transitions, the steady-state average density of the largest cluster $s_{{\rm max}}/L^2$ is typically considered the order parameter $\phi$. In the main text, the critical exponents $\beta,\gamma,\frac{\beta}{\nu}$ and $\frac{\gamma}{\nu}$ of the order parameter $\phi$ and its second cumulant $\chi$ were determined using Monte Carlo simulations and are listed in Table \ref{tab:table1}. We further determine these exponents $\beta, \gamma, \nu$ by analyzing the finite-size scaling properties of the order parameter and its second and fourth cumulant ({\it i.e.} the susceptibility $\chi$ and the Binder cumulant $B$ respectively). For  the PCs,  we   employ  the  finite size scaling analysis,
 \begin{eqnarray}
 \label{eq:sup_fss}  
 B = f_{b}(\Delta L^{\frac1\nu});~
 \phi = L^{-\frac{\beta}{\nu}}f_\phi(\Delta L^{\frac1\nu}); ~ 
 \chi = L^{\frac\gamma\nu}f_\chi(\Delta L^{\frac1\nu})
\end{eqnarray}
using  $\eps^*$ as  the   control parameter and  $\Delta = \eps^* - \eps_c;$  $\eps_c$ is the critical  energy threshold  which corresponds  to a  definite average density $\rho(\eps_c).$ For $\lambda = -0.1,0.1,$ the data collapses are presented in Fig. \ref{fig:collapse_lm-0.1_0.1} of the main text. In this supplemental material, we extend our analysis to additional values of $\lambda=-0.2,0,0.05,0.15,$ and $0.2$ and verify their consistency with the exponents obtained in Fig. \ref{fig:exponents}. In each of the figures, the data collapses for $B$ vs $\Delta L^{1/\nu}$ for different values of $\lambda$ are shown in Figs. \ref{fig:collapse_lm-0.2} - \ref{fig:collapse_lm0.2} (a), whereas Figs. \ref{fig:collapse_lm-0.2} - \ref{fig:collapse_lm0.2} (b), (c) describe the scaling collapse for $\phi$ vs $\Delta L^{1/\nu}$ and $\chi$ vs $\Delta L^{1/\nu}$ respectively for different values of $\lambda$.
\vspace{0.5cm}
\vspace{0.5cm}
\begin{figure}[h]
\centering
\includegraphics[width=5.6cm]{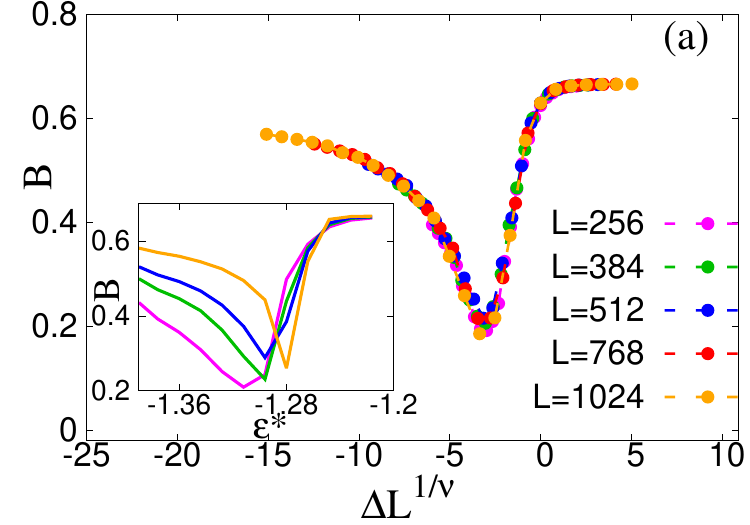}
\includegraphics[width=5.6cm]{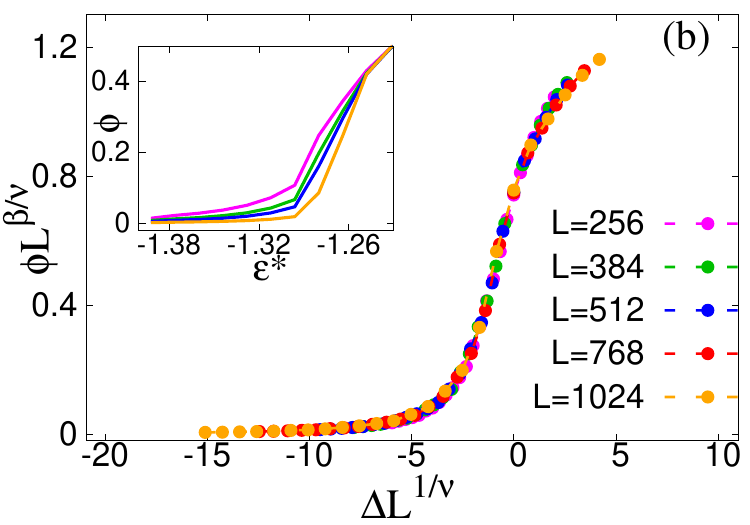}
\includegraphics[width=5.6cm]{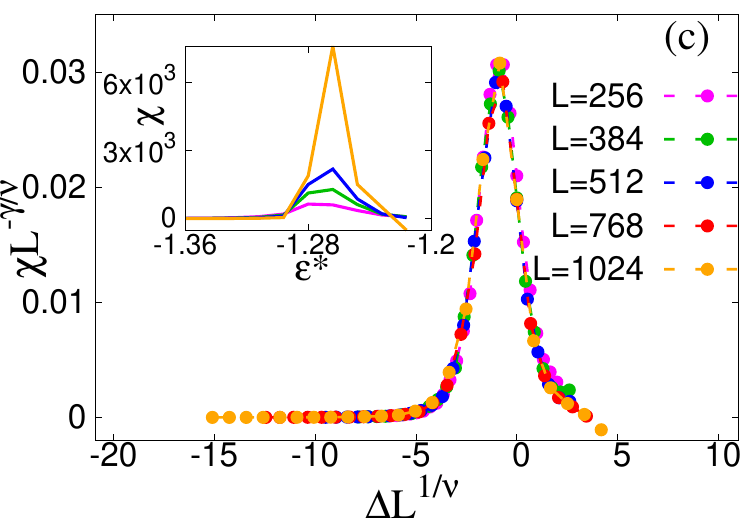}
\vspace*{-.2 cm}
\caption{ Data collapse of (a) Binder cumulant $B$, (b) $\phi L^{\beta/\nu}$ and (c) $\chi L^{-\gamma/\nu}$ as a function of $\Delta L^{1/\nu}$, across system sizes $L=256,384,512,768,1024$  to a unique scaling function observed for $\lambda = 0.0$. At the critical threshold $\epsilon^*_c=1.2525(2),$ the
best collapse is obtained for $ \nu=1.333, \beta = 0.138, \gamma=2.388.$ The uncollapsed plots are shown in the inset. Data are averaged over $10^6$ or more samples in a steady state.}\label{fig:collapse_lm0.05}
\end{figure}
\vspace{0.5cm}
\begin{figure}[h]
\centering
\includegraphics[width=5.6cm]{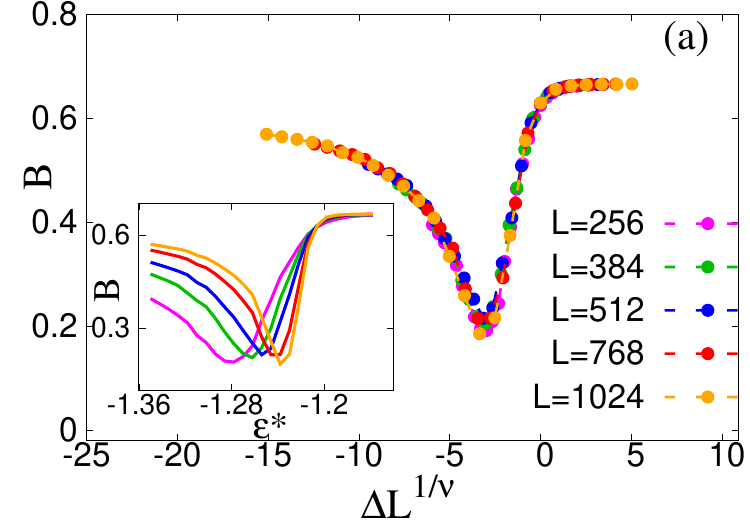}
\includegraphics[width=5.6cm]{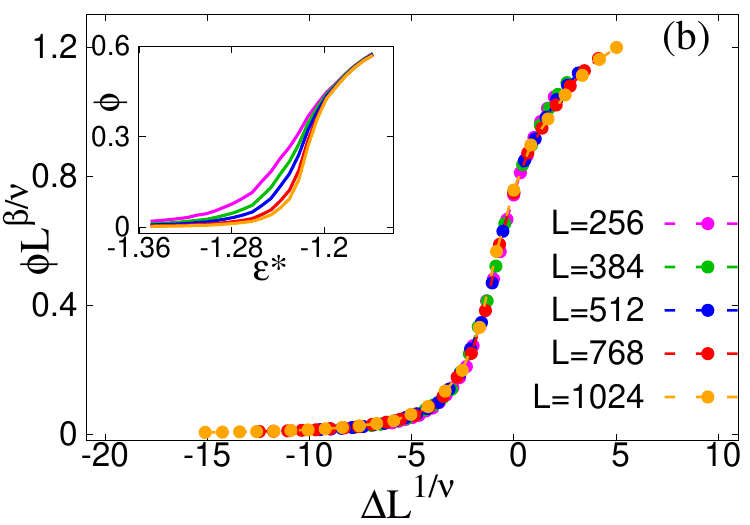}
\includegraphics[width=5.6cm]{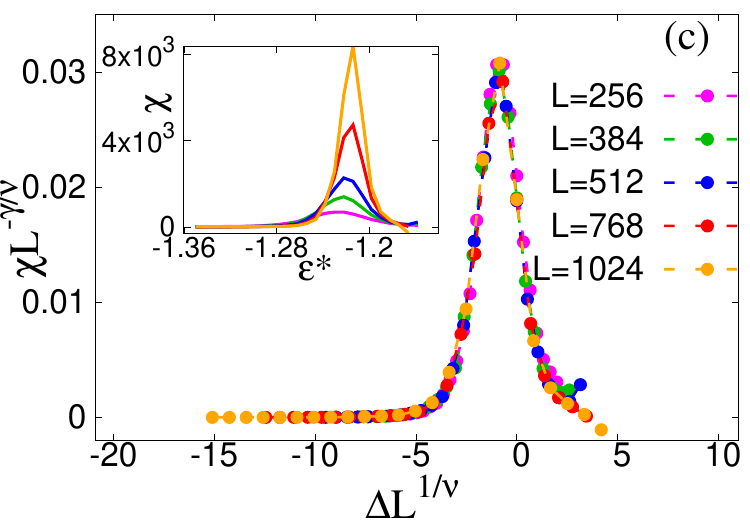}
\vspace*{-.2 cm}
\caption{ Data collapse of (a) Binder cumulant $B$, (b) $\phi L^{\beta/\nu}$ and (c) $\chi L^{-\gamma/\nu}$ as a function of $\Delta L^{1/\nu}$, across system sizes $L=256,384,512,768,1024$  to a unique scaling function observed for $\lambda = 0.05$. At the critical threshold $\epsilon^*_c=-1.2060(2),$ the
best collapse is obtained for $ \nu=1.434, \beta = 0.157, \gamma=2.571.$ The uncollapsed plots are shown in the inset. Data are averaged over $10^6$ or more samples in a steady state.}\label{fig:collapse_lm0.05}
\end{figure}
\vspace{0.5cm}
\vspace{0.5cm}

\end{document}